\documentclass[%
    reprint,
    superscriptaddress,
    nofootinbib,
    amsmath,amssymb,
    aps,
]{revtex4-2}

\usepackage[]{graphicx}% Include figure files
\usepackage{dcolumn}% Align table columns on decimal point
\usepackage{bm}% bold math
\usepackage[colorlinks=true, allcolors=blue]{hyperref}
\usepackage{booktabs}
\usepackage{subcaption}
\usepackage{cancel}
\usepackage{braket}
\usepackage{multirow, tabularx}
\usepackage{amsmath, amsthm, amssymb}
\usepackage{comment}
\usepackage[toc,page]{appendix}
\usepackage[dvipsnames]{xcolor}
\usepackage{ulem}
\usepackage{svg}

\usepackage{dsfont}
\usepackage{xcolor}
\usepackage{todonotes}
\usepackage{ulem}
\usepackage{listings}

\begin{document}

\title{A Sagnac-based arbitrary time-bin state encoder for quantum communication applications}

\author{Kannan~Vijayadharan}
\thanks{These two authors contributed equally}
\affiliation{Dipartimento di Ingegneria dell'Informazione, Universit\`a degli Studi di Padova, Via Gradenigo 6B - 35131 Padova, Italy.}

\author{Matías R.~Bolaños}
\thanks{These two authors contributed equally}
\affiliation{Dipartimento di Ingegneria dell'Informazione, Universit\`a degli Studi di Padova, Via Gradenigo 6B - 35131 Padova, Italy.}

\author{Marco~Avesani}
    \affiliation{Dipartimento di Ingegneria dell'Informazione, Universit\`a degli Studi di Padova, Via Gradenigo 6B - 35131 Padova, Italy.}
    \affiliation{Padua Quantum Technologies Research Center, Universit\`a degli Studi di Padova, via Gradenigo 6B, IT-35131 Padova, Italy}
    
\author{Giuseppe~Vallone}
\affiliation{Dipartimento di Ingegneria dell'Informazione, Universit\`a degli Studi di Padova, Via Gradenigo 6B - 35131 Padova, Italy.}
\affiliation{Padua Quantum Technologies Research Center, Universit\`a degli Studi di Padova, via Gradenigo 6B, IT-35131 Padova, Italy}

\author{Paolo~Villoresi}
\affiliation{Dipartimento di Ingegneria dell'Informazione, Universit\`a degli Studi di Padova, Via Gradenigo 6B - 35131 Padova, Italy.}
\affiliation{Padua Quantum Technologies Research Center, Universit\`a degli Studi di Padova, via Gradenigo 6B, IT-35131 Padova, Italy}

\author{Costantino~Agnesi}
    \email{costantino.agnesi@unipd.it}
    \affiliation{Dipartimento di Ingegneria dell'Informazione, Universit\`a degli Studi di Padova, Via Gradenigo 6B - 35131 Padova, Italy.}
    \affiliation{Padua Quantum Technologies Research Center, Universit\`a degli Studi di Padova, via Gradenigo 6B, IT-35131 Padova, Italy}

\begin{abstract}
Time-bin encoding of quantum information is highly advantageous for long-distance quantum communication protocols over optical fibres due to its inherent robustness in the channel and the possibility of generating high-dimensional quantum states. 
The most common implementation of time-bin quantum states using unbalanced interferometers presents challenges in terms of stability and flexibility of operation. 
In particular, a limited number of states can be generated without modifying the optical scheme. 
Here we present the implementation of a fully controllable arbitrary time-bin quantum state encoder, which is easily scalable to arbitrary dimensions and time-bin widths. 
The encoder presents high stability and low quantum bit error rate ($\text{QBER}$), even at high speeds of operation. 
Additionally, we demonstrate phase randomization and phase encoding without additional resources.
\end{abstract}

\maketitle

\section{Introduction}

The realization of a large-scale quantum network requires the transfer of quantum information and the generation of entanglement between the end nodes. 
In this regard, different photonic degrees of freedom such as polarization, frequency, transverse spatial modes, or time bins are utilized to encode and transmit quantum information \cite{Pirandola2020}. 
The choice of the photonic degree of freedom is highly dependent on the characteristics of the quantum channel employed to connect the different nodes. 
For example, optical fibres are often used for metropolitan and sub-urban links. 
This allows us to scale quantum networks using existing telecommunication infrastructure~\cite{PhysRevLett.121.190502,dynes2019cambridge,ribezzo2023deploying,martin2024madqci}. 
Whereas polarization encoding has been successfully used for Quantum Key Distribution (QKD) in such scenarios~\cite{Avesani:21,Telebit}, the presence of aerial fibres can render its implementation quite cumbersome~\cite{Li:18}. 
For this reason, time-bin is a common choice for such scenarios. Furthermore, time-bin encoding can support higher-dimensional protocols~\cite{doi:10.1126/sciadv.1701491, PhysRevX.9.041042}, which allow higher channel capacities and can be advantageous in particular conditions, such as with highly noisy channels~\cite{PhysRevA.82.030301}. 
From a more fundamental perspective, high-dimensional entanglement exhibits stronger non-classical correlations \cite{PhysRevLett.85.4418}.

Time-bin qubits are created by encoding quantum information in the arrival time of the photon and in their relative phases. 
Typically, this is achieved by using a pulsed laser and an unbalanced interferometer \cite{PhysRevLett.121.190502, PhysRevA.66.062308} or by carving pulses from a continuous wave laser using an intensity modulator driven by electrical signals \cite{Takesue:09, doi:10.1126/sciadv.1701491, vagniluca2020efficient}. 
A time-bin state encoder requires a number of features depending on the intended use cases. 

For QKD applications, for example, precise state encoding is required, while also maintaining a controlled phase relation between the early and late states. 
Moreover, different receivers might require different time-bin separations. 
When using the unbalanced interferometer approach, achieving the first point implies the addition of extra intensity modulators to both encode states and implement decoy state protocols, as well as extra phase modulators for phase encoding. 
Not only that, but the requirement for a Mach-Zehnder interferometer (MZI) both at the source and receiver sides makes the phase stabilization process considerably more challenging, which can in turn have detrimental effects on the quantum bit error rate (QBER) over time. 
Finally, modifying the time-bin separation proves difficult due to optical delay lines allowing only for fine tuning. 
On the other hand, the pulse carving approach exploit MZIs that require dedicated stabilization modules for their correct functioning. While such modules can be purchased or developed in-house~\cite{Iskander2019}, the use of these devices increases the overall complexity and cost of the system. 
While recent work by Kim et al. \cite{Kim2024} demonstrate the creation of a time-bin qubit with full control over the probability amplitudes, for complete generality the dimensionality also needs to be controlled.
When using the carving approach for QKD, phase randomization has to be carefully implemented, since non-randomized qubits can lead to quantum hacking attacks~\cite{PhysRevA.88.022308}, however, not all proof-of-principle QKD experiments take this into account~\cite{ribezzo2023deploying}.

For entanglement generation applications, a stable phase relation between the early and late state is also required, while several applications such as the violation of Bell inequalities beyond CHSH require the creation of a non-maximally entangled state \cite{PhysRevLett.108.100402}. 
Using only an unbalanced interferometer allows for the creation of only maximally entangled states such as the Bell state $\ket{\Phi^+}$, unless extra intensity and phase modulators are added. Moreover, as in the case for QKD, pulse carving approaches also require MZI's stabilization schemes.

In this paper, we propose the architecture for an arbitrary time-bin quantum state encoder using commercial off-the-shelf fibre components. 
Our scheme provides all the advantages of the carving approach, while also tackling the issue of active stabilization for Mach-Zehnder modulators by exploiting a modulation scheme based on Sagnac interferometers and taking advantage of its intrinsic stability~\cite{Roberts:18, iPognac}. 
We then provide experimental results and performance analysis of the proposed encoder, highlighting its capabilities of generating arbitrary time-bin states.

\section{Encoder design}

\begin{figure*}
    \centering
    \includegraphics[width=\linewidth]{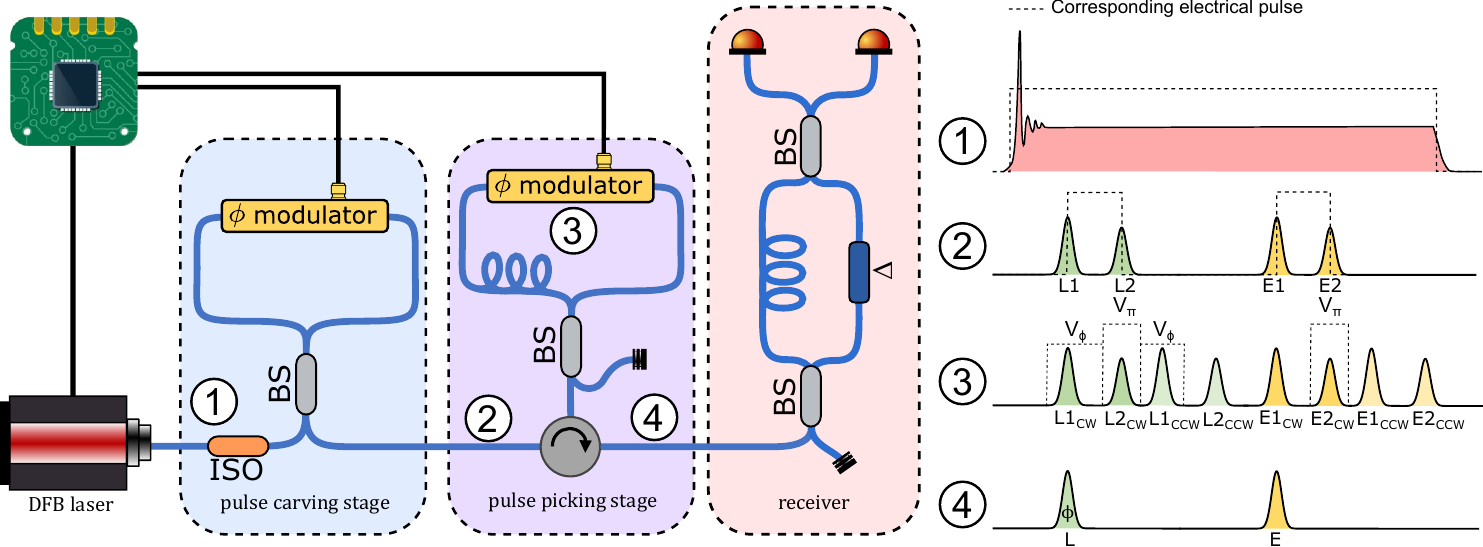}
    \caption{Experimental design (left) and qualitative explanation of its working principle (right). \textbf{(1)} In a first stage, if phase randomization between subsequent time-bin states is required, the laser can be gain-switched past its settling time until it reaches the steady state. If instead phase randomization is not required, the laser is switched on in continuous wave mode. \textbf{(2)} The laser output is sent through a symmetric Sagnac interferometer with an electro-optical modulator (EOM) from which two optical pulses are created for every electrical one. \textbf{(3)} Each pair of pulses is sent through another Sagnac interferometer where one of the two is extinguished, while an arbitrary phase can be applied to the non-extinguished one. \textbf{(4)} The final output of the encoder is now an arbitrary time-bin state with phase encoding. ISO: Isolator, BS: beamsplitter, $\phi-$modulator: phase modulator, $\Delta$: fibre stretcher.}
    \label{fig:encoder-design}
\end{figure*}

We propose a time-bin state encoder consisting of three components: a distributed feedback (DFB) laser and two Sagnac interferometer-based intensity-phase modulation stages (Fig. \ref{fig:encoder-design}). 
In the first step, if phase randomization is a requirement, we gain-switch a laser by sending sufficiently long electrical current pulses such that it reaches a steady-state emission~\cite{Paraso2021}, from where pulses will be carved (See Fig. \ref{fig:encoder-design}(1)). 
By gain-switching this way and carving pulses from the steady state, we guarantee that the phase relation between carved pulses is constant, thus creating a coherent superposition between them, while also having phase randomization between different pairs of gain-switched pulses, a necessary requirement for QKD applications. 

In the first intensity modulation stage, the \textit{pulse carving stage}, we use a Sagnac-based intensity modulator in the regime in which the width of the electrical pulse ($\Delta t_{\text{carving}}$) is much larger than the asymmetry in the Sagnac interferometer ($\delta t_{\text{asymm}}$), which for our case corresponds to the intrinsic asymmetries due to the length of the LiNbO$_3$ waveguide in the electro-optical phase modulator (EOM). 
In this regime, the output interferes constructively only when there is a phase difference $\phi$ between the CW (clockwise) and CCW (counter-clockwise) components, thus creating two optical pulses of width $\delta t_{\text{asymm}}$ separated by $\Delta t_{\text{carving}}$ and with amplitude $\propto\sin^2(\phi)$, as represented in Fig. \ref{fig:encoder-design}(2).
Although this stage alone could be used to create a 2-dimensional time-bin state, discrepancies in the rising and falling edge of the electrical pulse would be reflected on the shape of the generated optical pulses, increasing distinguishability between the two. 

To remove the distinguishability from the pulse carving stage, we added a second intensity modulation stage, the \textit{pulse picking stage}, in which one of the two optical pulses generated by a single electrical pulse is extinguished. 
To do so, we use an intentionally unbalanced Sagnac interferometer by adding a fixed-length optical fibre (equivalent to an optical delay of $\delta t_{\text{asymm}_2}$) to one of the branches, after which we apply a $\pi$ phase shift to either the CW or the CCW component, as shown in Fig. \ref{fig:encoder-design}(3) \cite{Roberts:18}. 
Following a similar logic to the pulse carving stage, since the output pulse amplitude is $\propto \sin^2(\phi)$, by applying a relative phase shift of $\pi$, the pulse is effectively extinguished \cite{Roberts:18}.
With both stages, a \textit{d} dimensional time-bin state can be created by sending $d$ pulses of width $\Delta t_{\text{carving}}$ separated by $\Delta t$ to the pulse carving stage, followed by extinguishing one of the two optical pulses created by each electrical pulse in the pulse picking stage, as shown in Fig. \ref{fig:encoder-design}(4).
Assuming a train of rectangular pulses in the pulse carving stage defined by
\begin{equation}
    V(t) = \sum_i^d V_i\Pi\left( \frac{t - i\cdot \Delta t}{\Delta t_{\text{carving}}} -t_0\right),
\end{equation}
with $\Pi\left(\frac{t}{\Delta t}\right)$ a rectangular pulse of width $\Delta t$ and unit amplitude centred at $t=0$, each pulse will create two optical pulses of width $\delta t_{\text{asymm}}$ separated by $\Delta t_{\rm carving}$ with amplitude $\propto sin^2\left(\frac{V_i}{V_\pi} \pi\right)$, where $V_\pi$ corresponds to the voltage necessary to apply a phase shift of $\pi$ on the electro-optical modulator. 
Then, after pulse-picking one of the two generated optical pulses per electrical pulse sent, the resulting quantum state corresponds to
\begin{equation}
    \ket{\Psi} = \sum_i^d \alpha_i \ket{t_0 + i\cdot \Delta t} = \sum_i^d \alpha_i \ket{t_i},
\end{equation}
where the state $\ket{t_i}$ corresponds to a weak coherent state centred at time $t_i$ and $\alpha_i$ is such that $\alpha^2_i=\frac{a^2_i}{\sum_ja^2_j}$, where $a^2_i=\sin^2\left(\frac{V_i}{V_\pi} \pi\right)$. Moreover, we have that $\sum_i a^2_i \propto \langle n \rangle$, with $\langle n \rangle $ the average number of photons of the output state, allowing the encoding of different mean number of photons by controlling the values of $V_i$.
Lastly, by applying $V_\phi$ to both the CW and the CCW components of the non-extinguished pulse on the pulse picking stage, the resulting quantum state becomes
\begin{equation}\label{eq:general-tbin}
    \ket{\Psi} = \sum_i^d e^{-i\phi_i} \alpha_i \ket{t_i},
\end{equation}
making our design capable of creating any arbitrary time-bin encoded qudit, including arbitrary dimension, time-bin separation, and phase encoding, all while requiring only two phase modulator. 

Moreover, for QKD applications where decoy state protocols are required, whereas current systems have to include extra intensity modulators, our encoder is capable of adding decoy states by tuning the value of $\sum_i a^2_i\propto\langle n \rangle$ on the pulse carving stage.

\begin{figure}
    \centering
    \includegraphics[width=\linewidth]{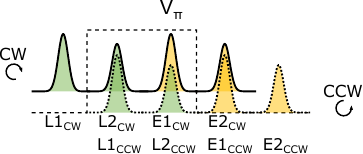}
    \caption{Pulse picking scheme to create the $\ket{-}$ state at higher repetition rate. This figure is analogous to the one represented in Fig. \ref{fig:encoder-design}(3).}
    \label{fig:pulse-picking-minus}
\end{figure}

It is worth noting that this scheme works for arbitrary states, as long as electrical pulses can be applied to a single optical pulse without affecting the rest. 
This condition is easily applied when 
\begin{equation}\label{asymm_condition_1}
\delta t_{\text{asymm}_2} > \Delta t + \Delta t_{\text{carving}}
\end{equation}
(i.e. all counter-clockwise components are after the clockwise ones) or
\begin{equation}\label{asymm_condition_2}
\delta t_{\text{asymm}_2}+\Delta t_{\text{carving}} < \Delta t
\end{equation}
(i.e. the counter-clockwise component of the early pulses are between the clockwise component of the early and late pulses), both of which give an upper bound on the achievable repetition rate for the states. 
Nevertheless, for QKD applications it is enough to create three states to obtain a secure BB84 protocol \cite{rusca2018security}, which is possible with a slightly different hardware condition in which the second stage asymmetry 
\begin{equation}
    \delta t_{\text{asymm}_2}\sim\delta t_{\text{asymm}}\sim\Delta t_{\text{carving}}.
\end{equation}
In this way, it is possible to create the $\ket{-}$ state simply by applying a single pulse during the pulse-picking stage (Fig. \ref{fig:pulse-picking-minus}). 
This method allows the three-state encoding for repetition rates $R<\frac{1}{5\delta t_{\text{asymm}}}$, while otherwise it would be upper bounded by $R<\frac{1}{\Delta t + \Delta t_{\text{carving}} + \delta t_{\text{asymm}_2}+ \delta t_{\text{asymm}}}$.
Moreover, if phase encoding is not required, it is possible to take the destructive port on the pulse picking stage to relax the $V_\pi$ requirement for the EOM while also reducing the intrinsic $\text{QBER}$ of the system.

\subsection*{Experimental implementation}

The encoder was built for wavelengths of 1550 nm and 780 nm, two relevant wavelengths in several quantum communication applications, particularly for QKD implementations in both free-space and fibre channels, since telecommunication wavelengths allows the use of already-deployed fibre in the metropolitan area, while 780 nm is useful for both free-space satellite channels, and to generate telecommunications wavelength entangled photons \cite{lu2022micius, li2025microsatellite}. 
For both 1550 nm and 780 nm encoder, we used a DFB laser as source
and two electro-optical phase modulators inside the Sagnac loops for both modulation stages.
For both the cases, to feed the electrical signals, we used a 
FPGA UltraScale+ development board by Xilinx, from which we used its balun-free DAC channels, each coupled to an RF amplifier placed before each EOM.

\section{Performance analysis}

The performance of the proposed time-bin encoder scheme is characterized by a series of tests, including both time-of-arrival measurements with the detector placed directly after the encoder output and interferometric measurements with time-bin receivers. 
From these measurements, it is possible to retrieve the relevant information on a time-bin qubit. 
From time-of-arrival measurements the probability amplitudes of any $d$-dimensional state, i.e. $|\alpha_i|^2$ from equation \ref{eq:general-tbin}, can be directly measured. 
On the other hand, by using the Franson interferometer, which is an unbalanced Mach-Zehnder interferometer (uMZI), and controlling the relative phase between both arms of the interferometer, it is possible to measure the relative phases on the state, i.e. $\phi_i$ from equation \ref{eq:general-tbin}. 
As a first instance, to control the relative phase in the receiver, a software-controlled piezoelectric fibre stretcher was used (see receiver on Fig. \ref{fig:encoder-design}).

\subsection{Phase stability}

\begin{figure}
    \centering
    \includegraphics[width=\linewidth]{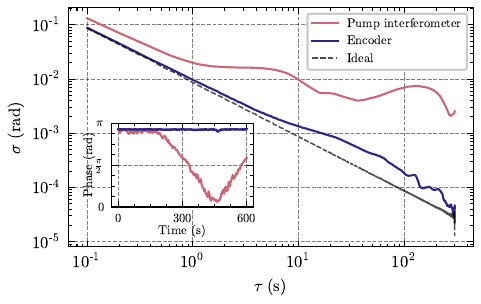}
    \caption{Allan deviation plot obtained at the receiver for a time-bin state generated with our encoder (in blue) and with an unbalanced interferometer (in red), both compared with the deviation obtained for a random walk signal sampled every 100 ms with $5$~mrad phase diffusion. Inset: Measured phase as a function of time from which the Allan deviation was obtained.
}
    \label{fig:phase-stability}
\end{figure}

The phase stability of the encoder is characterized by measuring the phase drift of the generated time-bin superposition over an extended measurement period. 
To do this, we first create a time-bin superposition state (the $\ket{-}$ state in this case) using the encoder and compare the phase drift against the same state created using an uMZI. 
For a fair comparison, in the latter case, a single pulse carved by the encoder is used to create the superposition state using the uMZI.

{Due to phase instabilities, the measured state at the receiver
is
\begin{equation}
    \ket{\psi}_{j} = \frac{1}{\sqrt{2}}(\ket{E}-e^{i\phi_s}\ket{L}),
\end{equation}
where $\phi_s$ is the phase drift of the source (ideally $\phi_s=0$).
}
The phase $\phi_s$ is estimated from the receiver measurements for both cases, and the phase stability is estimated. 
Even though the measurements for both cases were taken at different times, they were done one immediately after the other under the same conditions. 
For both cases, the uMZIs in the receiver side were properly passively stabilized in a closed isolated container and left to stabilize for $\sim 1$ hour. 
To quantify the phase stability over time, the Allan deviation $\sigma$, defined such that

\begin{equation}
\sigma^2(\tau)=\frac{1}{2 \tau^2}\left\langle\left(x_{n+2}-2 x_{n+1}+x_n\right)^2\right\rangle
\end{equation}
where $x_i$ is the $i$-th of the phase values spaced by the measurement interval $\tau$ (Fig. \ref{fig:phase-stability}). The results show significant phase stability using the encoder, highlighting its advantage over an unbalanced interferometer to create time-bin states. 
It is worth noting that the unbalanced interferometer at the source could be further stabilized to improve its stability over time, but doing so implies extra resources and complexity of the system \cite{vsvarc2023sub}. The purpose of this characterization is to highlight the fact that, when little to no stabilization scheme is used on the source side (be it with the encoder or uMZI), the encoder presents higher intrinsic stability.

\subsection{QKD Experiment}

To demonstrate the capabilities of our encoder while highlighting its versatility for different time-bin receivers, we used the proposed time-bin encoder to generate a fixed pseudo-random sequence of 512 symbols to simulate a BB84 QKD implementation with decoy state. 
{To simulate a QKD system as faithfully as possible, the system was prepared with typical values of decoy state BB84 experiments} \cite{rusca2018finite}.
With this in mind, we chose $\nu/\mu\sim0.3$, with $\nu=0.2$ and $\mu=0.6$ the mean number of photon for decoy and signal pulses respectively, and we generated the 512 symbol sequence with $p_Z=87\%$ and $p_X=13\%$ with $p_{Z(X)}$ the probability to generate a state on the key (check) basis.

\begin{figure}
    \centering
    \includegraphics[width=\linewidth]{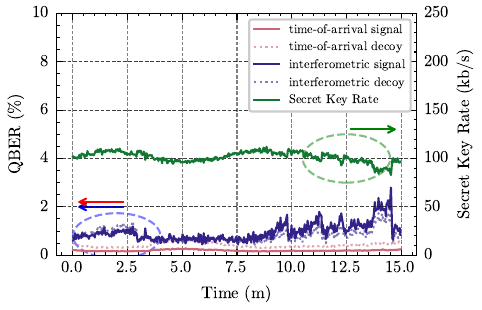}
    \caption{$\text{QBER}$ on the $Z$ and $X$ basis for a 512-symbol sequence over time. In green, associated with the right axis, the SKR obtained over time.}
    \label{fig:qber-512-long}
\end{figure}

\begin{figure}
    \centering
    \includegraphics{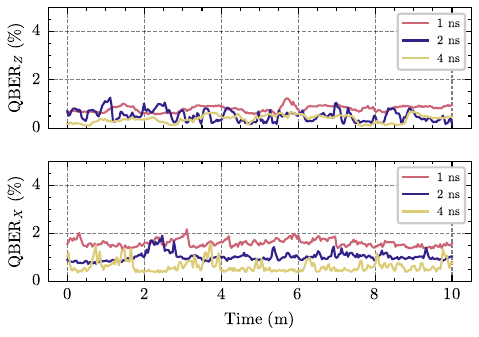}
    \caption{$\text{QBER}$ on the $Z$ and $X$ basis for a 512-symbol sequence over time for different time-bin $\Delta t$ with a fixed repetition rate $R=100$ MHz.
    }
    \label{fig:qber-100mhz}
\end{figure}

\begin{figure}
    \centering
    \includegraphics{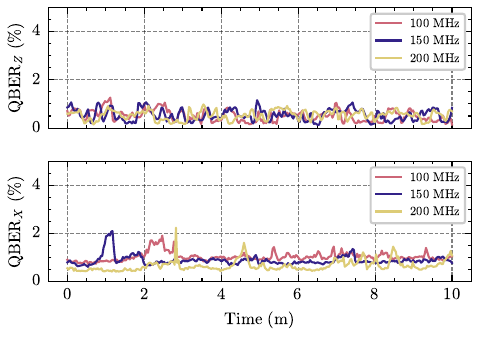}
    \caption{$\text{QBER}$ on the $Z$ and $X$ basis for a 512-symbol sequence over time for different repetition rate $R$ with a fixed time-bin $\Delta t=2$ ns.}
    \label{fig:qber-2ns}
\end{figure}

As a first instance, to demonstrate the best achievable performance for our encoder, we performed the 4-state version of the BB84 protocol at a fixed $100$ MHz repetition rate with a more limiting, but better performing, hardware configuration, corresponding to equation \ref{asymm_condition_1}.
We measured $\text{QBER}$ on both the time-of-arrival basis (i.e. the key basis) and on the superposition basis (i.e. the check basis) for an hour to evaluate the performance and long-term stability of the device (Fig. \ref{fig:qber-512-long}). From this measurement, the secret key rate was estimated along the entire duration, obtaining an average $\text{SKR} = 101.66\pm5.31$ kb/s at 25 dB of total losses. To estimate the secret key rate, decoy state bounds were applied, taking into account finite-key effects \cite{rusca2018finite}.
Moreover, by performing the measurement in a long pseudo-random sequence, we can evaluate possible cross-talk or patterning effects.
\begin{figure}
    \centering
    \includegraphics[width=\linewidth]{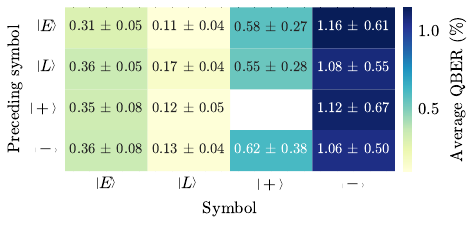}
    \caption{Average $\text{QBER}$ obtained for each symbol type for all possible preceding symbols. Due to low statistics of $\ket{+}$ and $\ket{-}$ symbols, combination of those two symbols are either not present or have considerably large relative error.} 
    \label{fig:qber-patterning}
\end{figure}
It was observed that the QBER values for both basis were kept under the 1\% threshold for most of its runtime, even with a non-fully-optimized active phase stabilization scheme on the receiver. 
By performing a first-neighbour patterning analysis (Fig. \ref{fig:qber-patterning}), we concluded that there is no complete distinguishability by changing the preceding symbol on the sequence. 
It is worth noting that the standard deviation of 0 for the $\ket{+},\ket{-}$ and  $\ket{-},\ket{+}$ sequences is due to that combination appearing only once over the 512-symbol sequence, while the $\ket{+},\ket{+}$ sequence does not appear at all, thus not having an associated $\text{QBER}$ value. 
Then, as a second instance, we performed a 3-state version of the BB84 protocol using both states of the key basis ($\ket{E}$ and $\ket{L}$) and only one state for the check basis ($\ket{-}$). 
The measurements were done with different repetition rates $R=100, 125, 200$ MHz for a fixed time-bin separation of $\Delta t = 2$ ns (Fig. \ref{fig:qber-100mhz}), and for different $\Delta t=1, 2, 4$ ns (and corresponding receivers) for a fixed repetition rate of $100$ MHz (Fig. \ref{fig:qber-2ns}).
We observed that on average the $\text{QBER}$ value was kept mostly under $1\%$ for the $Z$ basis, while under $2\%$ for the $X$ basis. 
By increasing $\Delta t$, we observed an overall improvement over the measured $\text{QBER}$, due to larger time-bins making alignment of the electrical signal less challenging. 
It was also observed that changing the repetition rate of the system does not significantly affect $\text{QBER}$.

\subsubsection{Phase randomization}

\begin{figure}
    \centering
    \includegraphics[width=\linewidth]{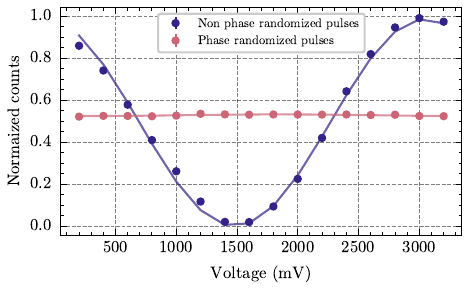}
    \caption{Phase scan applied to both phase randomized (in red) and non-phase randomized pulses when carving from a continuous wave laser (in blue).}
    \label{fig:phase-randomization}
\end{figure}

Phase randomization between consecutive time-bin states is a requirement for QKD applications, with a non-optimal phase randomization negatively affecting secret key generation. \cite{curras2023security}. The phase randomization of the time-bin states is obtained by carefully gain-switching the laser and fully depleting the cavity before carving the next state to ensure that the phase is uncorrelated with the preceding one. From a direct measurement of the gain-switched laser pulse, we estimated the settling time $t_{\rm ss}$ (i.e. the time required for the laser to reach the steady-state after gain-switching) and the off time $t_{\rm off}$ (i.e. the time required for the laser to fully deplete the cavity). These parameters provide an upper bound for the repetition rate of the system, such that $R<(t_{\rm ss}+t_{\rm off})^{-1}$, so we characterized it with the minimum $t_{\rm off}$ possible such that phase randomization happened. 

To characterize phase randomization using the same time-bin receiver as in the QKD experiment, the electrical pulses were carefully aligned such that the early pulse is carved close to the laser turning off, while the late pulse is carved from the consecutive laser pulse, guaranteeing that $\Delta t > t_{\rm ss} + t_{\rm off}$. On the other hand, to study the non-phase randomized scenario, we first carved the time-bin state from a single gain-switched laser pulse, and in a later stage carved it from a continuous wave laser.
We then verified the phase randomization of the pulses by performing interferometric measurements using the time-bin receiver and sending states at a repetition rate of 100 MHz.

\begin{figure}
    \centering
    \includegraphics[width=\columnwidth]{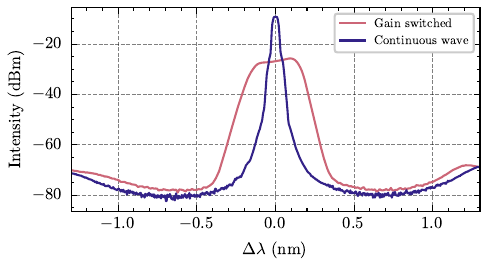}
    \caption{Comparison of the spectra of the {DFB} laser in gain-switched mode (in red) and continuous-wave mode (in blue). The x-axis is the relative shift around the central wavelength.}
    \label{fig:spectrum}
\end{figure}

{When a phase scan is performed on two phase randomized pulses by inducing a relative phase between them, there is a low visibility of the interference fringe $V_{pr}=0.7$\%, while the visibility of the non-phase randomized pulses when carving from a single gain-switched pulse was obtained to be $V_{npr}=92.4$\%. As this visibility value was considerably lower than expected, the same analysis was performed when carving the time-bin states from the same laser in continuous wave operation, for which the visibility was obtained to be $V_{npr}=98.7$\% (Fig. \ref{fig:phase-randomization}). We attributed this discrepancy in visibilities to the fact that the laser spectrum is modified when gain-switching, increasing the linewidth of the laser and as such reducing phase stability. To verify this behaviour, we obtained the spectrum of the DFB laser both with and without gain switching it using an AQ6370D Optical Spectrum analyser by Yokogawa (Fig. \ref{fig:spectrum}). Nevertheless, using our device's phase encoding scheme, with the addition of a Quantum Random Number Generator (QRNG), it is possible to perform phase randomization on the states without gain-switching techniques, and without requiring an extra phase modulator. Due to the embedded system nature of the {FPGA} development board, QRNG-based phase encoding to perform phase randomization is fairly straightforward to implement \cite{9695406, francesconi2024scalable}.}

As shown by Currás-Lorenzo et al. in \cite{curras2023security}, a non-zero degree of phase correlation reduces the maximum achievable key rate. Following the method in \cite{grunenfelder2020performance}, by estimating the visibility for both scenarios, we obtained a degree of phase correlation $p^*_c=0.008$, implying a \textit{source quality} $q=0.97$. We attributed this non-optimal source quality to the fact 
{that we set the $t_{\rm off}$ as short as possible, and we expect the source quality to improve for lower repetition rates and longer $t_{\rm off}$}.

\subsection{High-dimensional state encoding}\label{sec:arb_encoding}

\begin{figure}[!ht]
    \centering
    \includegraphics[width=\linewidth]{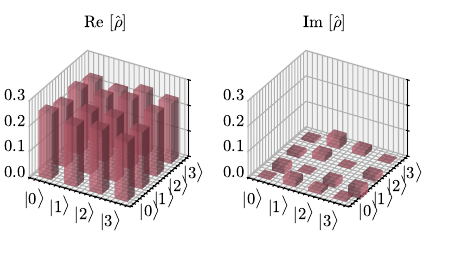}
    \caption{Reconstructed density matrix for $\ket{\psi^4_0}$.}
    \label{fig:density}
\end{figure}

To demonstrate the capability to create arbitrary high-dimensional states described by Eq. \ref{eq:general-tbin}, a sequence of four-dimensional time-bin states composed of
$$\ket{\psi^4_0} = \frac{\ket{t_0} + \ket{t_1} + \ket{t_2} + \ket{t_3}}{2}, $$
$$\ket{\psi^4_1} = \frac{\ket{t_0} + e^{-\pi i/3}\ket{t_1} + e^{-2\pi i/3}\ket{t_2} + e^{-\pi i}\ket{t_3}}{2},$$
was prepared and measured at a repetition rate of 40 MHz with $\Delta t=3$ ns.

To perform quantum state tomography, we measured the time-bin state using cascaded MZIs with delays of $\Delta t$ and $2\Delta t$, and phase differences $\phi_1$ and $\phi_2$. To reconstruct the complete density matrix, measurements corresponding to the 4-dimensional Pauli matrices $\sigma_X$, $\sigma_Y$ and $\sigma_Z$ are required \cite{PhysRevA.66.012303}.
By placing a 90:10 beamsplitter before the interferometer and measuring the direct time of arrival of the input state, the time-bin basis or $\sigma_Z$ can be measured. The interference measurements from the cascaded uMZI correspond to $\sigma_X$ or $\sigma_Y$ depending on the phases $\phi_1$ and $\phi_2$ being $0$ or $\frac{\pi}{2}$ respectively.

As demonstrated in \cite{Takesue:09}, only two detectors are required to perform a full state tomography on a four-dimensional time-bin state using cascaded MZIs. Additionally, in order to mitigate the experimental challenge of fluctuation of the photon count rate and phase drift between different measurements, we implemented a simultaneous measurement of the three bases using a single high-speed electro-optical phase modulator external to the uMZI and synchronised with the source. By applying a four-symbol phase sequence to the state at a rate one-fourth that of the source generation rate, the measurement settings are effectively changed between subsequent time-bins (see Appendix \ref{appendix:tomography}).

The four-symbol phase-sequence was engineered as to be able to reconstruct all relevant $\langle \sigma_i\rangle$ with $i=X,Y,Z$. The first symbol applies no phase to the state, allowing the extraction of $\langle \sigma_X\rangle$ from the interference fringes of the cascaded interferometers. Meanwhile, the following 3 symbols apply a $\phi=\pi/2$ phase over the pulses corresponding to $\ket{t_0}$ and $\ket{t_k}$ with $k=1,2,3$ respectively for each symbol. By measuring the amplitudes of a particular subset of the interference fringes, all expectation values $\langle \sigma_Y\rangle$ can be reconstructed.

From the detection probabilities corresponding to the measurement settings applied, the density matrix $\rho$ of the measured state was reconstructed by maximum likelihood estimation (Fig. \ref{fig:density}). The fidelity of the measured state was calculated by $\mathcal{F}=\braket{\psi_{ideal}|\rho|\psi_{ideal}}$ to be 0.969 with respect to $\ket{\psi^4_0}$. We attribute the infidelity to a slight mismatch in the applied phase $\phi$ from the desired value of $\pi/2$ and phase drift in the measurement interferometer, as is evident from the imaginary part of $\rho$ in Fig. \ref{fig:density}.

A different approach was used for characterising the measured state for $\ket{\psi^4_1}$. Instead of the complete set of 16 measurements required for the state tomography, a reduced subset was used by measuring just the overlap with the ideal state to calculate the fidelity as $\mathcal{F}=|\braket{\psi|\psi_{ideal}}|^2$ yielding $\mathcal{F}=0.989\pm0.011$. It is to be noted that in an adversarial scenario or without prior knowledge about the prepared state, a full tomography is required, since this assumes purity of the state.

\subsection{Time-bin entanglement}
A time-bin entangled state is created by pumping a nonlinear crystal with a time-bin qubit to generate a pair of entangled photons by spontaneous parametric downconversion (SPDC) process.

\begin{figure}
    \centering
    \includegraphics[width=\linewidth]{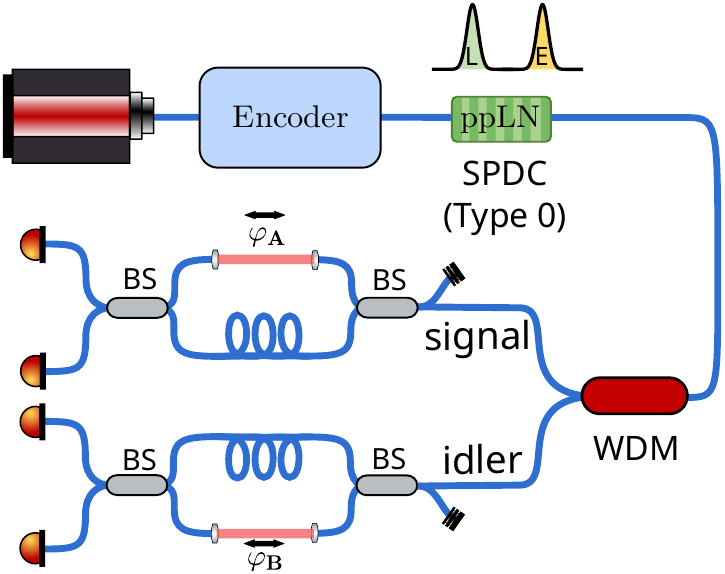}
    \caption{Simplified experimental design for the time-bin entanglement scheme. WDM: Wavelength Division Multiplexing filter, BS: beamsplitter.}
    \label{fig:ent-scheme}
\end{figure}

Here, we start with an equal superposition state created by the encoder as the pump photon with the state
\begin{equation}
    \ket{\psi}_{pump} = \frac{1}{\sqrt{2}}(\ket{E}+e^{i\phi}\ket{L}),
\end{equation}
to then generate the maximally entangled $\ket{\Phi^+}$ state (Fig. \ref{fig:ent-scheme}).
The repetition rate and bin separation were 200 MHz and 2 ns respectively. A periodically poled lithium niobate waveguide is used for Type-0 SPDC process, producing the state
\begin{equation}
    \ket{\psi}_{PDC} = \ket{\Phi^+} = \frac{1}{\sqrt{2}}(\ket{E}_1\ket{E}_2+e^{i\phi}\ket{L}_1\ket{L}_2).
\end{equation}

\begin{figure}
    \centering
    \includegraphics[width=\linewidth]{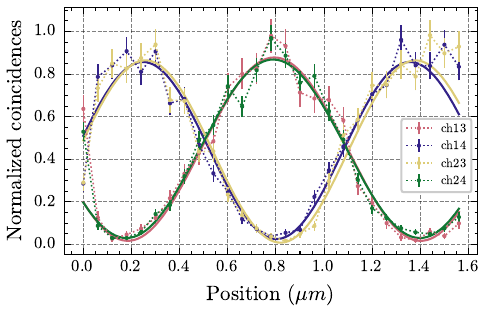}
    \caption{Modulation of the rate of postselected coincidences between the signal and idler photons with respect to the relative phase introduced in the receiver interferometer. Measured data from the four detector channels plotted along with a cosine fit.}
    \label{fig:entanglement}
\end{figure}

The entangled photon pairs are separated by wavelength filtering and sent to the two state analysers to perform projective measurements on the state. 
The state analysers are identical unbalanced interferometers with a delay equal to the bin separation of 2 ns. 
By post-selecting on the middle peak in the detection histogram, measurements are performed on the two photons in the superposition basis with the coincidence probability given by $\frac{1}{4}(1+\cos{(\phi_A+\phi_B)})$, where $\phi_A$ and $\phi_B$ correspond to the local phase shift introduced by each state analyser. 
By performing a phase scan on one of the receiver interferometers using a nanometric translation stage, we observe a visibility of up to 96\% in the coincidence count rate (Fig. \ref{fig:entanglement}), certifying the entanglement and corresponding to a CHSH violation with $S=2.72\pm0.037$.

For certifying the entanglement, only the detection events exhibiting quantum interference were postselected. 
This opens up the postselection loophole in Bell tests \cite{Jogenfors_2014}, which can be circumvented by using fast optical switches \cite{PhysRevLett.121.190401} or specialized interferometric geometries \cite{Santagiustina:24}, but this is beyond the scope of this paper. 
Furthermore, while we have only shown the scenario of a maximally entangled state, the encoder is capable of creating non-maximally entangled states as well using unequal arbitrary superposition states to pump the downconversion process.
However, measurement of such states requires more complex receiver architectures that involve high-speed optical switches and reconfigurable beamsplitters.

\section{Conclusions}
We have demonstrated a novel arbitrary time-bin state encoder with a low bit error rate and high long-term stability. Our device shows higher stability over time compared with an unbalanced interferometer to create time-bin states, even without any active stabilization methods. 
The pulses carved from a continuous-wave laser allowed for higher spectral purity and low chirp operation compared to directly modulated lasers, and tunability in pulse duration compared to mode-locked lasers. 
The inherent stability of the Saganac interferometer also leads to stable long-term operation that mitigates bias drift problem in Mach-Zehnder modulator-based approaches for carving pulses.

We have shown the capability of the encoder to create arbitrary time-bin-encoded qudits, including phase encoding, without increasing the design complexity and number of optical components. 
The ability to encode relative phases on the time-bin state would allow, by including an extra QRNG, to phase-randomise subsequent states without affecting the source spectrum, as would happen for gain-switching regimes.
The capability to not affect the spectrum has a direct effect, for example, on the quantum bit error rate (QBER) of the source for QKD applications, as it depends on the interference visibility, which we have shown is lower for gain-switching regimes, implying a higher intrinsic QBER. 

We showed the applicability of the encoder for quantum communication applications, in particular for QKD and entanglement generation, with reconfigurability of parameters while maintaining a low error rate. 
The arbitrary time bin state preparation we have demonstrated here allows for adaptability to existing/deployed time-bin quantum communication systems.

\bibliography{ref} % Produces the bibliography via BibTeX.

\begin{thebibliography}{35}%
\makeatletter
\providecommand \@ifxundefined [1]{%
 \@ifx{#1\undefined}
}%
\providecommand \@ifnum [1]{%
 \ifnum #1\expandafter \@firstoftwo
 \else \expandafter \@secondoftwo
 \fi
}%
\providecommand \@ifx [1]{%
 \ifx #1\expandafter \@firstoftwo
 \else \expandafter \@secondoftwo
 \fi
}%
\providecommand \natexlab [1]{#1}%
\providecommand \enquote  [1]{``#1''}%
\providecommand \bibnamefont  [1]{#1}%
\providecommand \bibfnamefont [1]{#1}%
\providecommand \citenamefont [1]{#1}%
\providecommand \href@noop [0]{\@secondoftwo}%
\providecommand \href [0]{\begingroup \@sanitize@url \@href}%
\providecommand \@href[1]{\@@startlink{#1}\@@href}%
\providecommand \@@href[1]{\endgroup#1\@@endlink}%
\providecommand \@sanitize@url [0]{\catcode `\\12\catcode `\$12\catcode `\&12\catcode `\#12\catcode `\^12\catcode `\_12\catcode `\%12\relax}%
\providecommand \@@startlink[1]{}%
\providecommand \@@endlink[0]{}%
\providecommand \url  [0]{\begingroup\@sanitize@url \@url }%
\providecommand \@url [1]{\endgroup\@href {#1}{\urlprefix }}%
\providecommand \urlprefix  [0]{URL }%
\providecommand \Eprint [0]{\href }%
\providecommand \doibase [0]{https://doi.org/}%
\providecommand \selectlanguage [0]{\@gobble}%
\providecommand \bibinfo  [0]{\@secondoftwo}%
\providecommand \bibfield  [0]{\@secondoftwo}%
\providecommand \translation [1]{[#1]}%
\providecommand \BibitemOpen [0]{}%
\providecommand \bibitemStop [0]{}%
\providecommand \bibitemNoStop [0]{.\EOS\space}%
\providecommand \EOS [0]{\spacefactor3000\relax}%
\providecommand \BibitemShut  [1]{\csname bibitem#1\endcsname}%
\let\auto@bib@innerbib\@empty
%</preamble>
\bibitem [{\citenamefont {Pirandola}\ \emph {et~al.}(2020)\citenamefont {Pirandola}, \citenamefont {Andersen}, \citenamefont {Banchi}, \citenamefont {Berta}, \citenamefont {Bunandar}, \citenamefont {Colbeck}, \citenamefont {Englund}, \citenamefont {Gehring}, \citenamefont {Lupo}, \citenamefont {Ottaviani}, \citenamefont {Pereira}, \citenamefont {Razavi}, \citenamefont {Shaari}, \citenamefont {Tomamichel}, \citenamefont {Usenko}, \citenamefont {Vallone}, \citenamefont {Villoresi},\ and\ \citenamefont {Wallden}}]{Pirandola2020}%
  \BibitemOpen
  \bibfield  {author} {\bibinfo {author} {\bibfnamefont {S.}~\bibnamefont {Pirandola}}, \bibinfo {author} {\bibfnamefont {U.~L.}\ \bibnamefont {Andersen}}, \bibinfo {author} {\bibfnamefont {L.}~\bibnamefont {Banchi}}, \bibinfo {author} {\bibfnamefont {M.}~\bibnamefont {Berta}}, \bibinfo {author} {\bibfnamefont {D.}~\bibnamefont {Bunandar}}, \bibinfo {author} {\bibfnamefont {R.}~\bibnamefont {Colbeck}}, \bibinfo {author} {\bibfnamefont {D.}~\bibnamefont {Englund}}, \bibinfo {author} {\bibfnamefont {T.}~\bibnamefont {Gehring}}, \bibinfo {author} {\bibfnamefont {C.}~\bibnamefont {Lupo}}, \bibinfo {author} {\bibfnamefont {C.}~\bibnamefont {Ottaviani}}, \bibinfo {author} {\bibfnamefont {J.~L.}\ \bibnamefont {Pereira}}, \bibinfo {author} {\bibfnamefont {M.}~\bibnamefont {Razavi}}, \bibinfo {author} {\bibfnamefont {J.~S.}\ \bibnamefont {Shaari}}, \bibinfo {author} {\bibfnamefont {M.}~\bibnamefont {Tomamichel}}, \bibinfo {author} {\bibfnamefont {V.~C.}\ \bibnamefont {Usenko}}, \bibinfo {author} {\bibfnamefont
  {G.}~\bibnamefont {Vallone}}, \bibinfo {author} {\bibfnamefont {P.}~\bibnamefont {Villoresi}},\ and\ \bibinfo {author} {\bibfnamefont {P.}~\bibnamefont {Wallden}},\ }\bibfield  {title} {\bibinfo {title} {Advances in quantum cryptography},\ }\href {https://doi.org/10.1364/AOP.361502} {\bibfield  {journal} {\bibinfo  {journal} {Adv. Opt. Photon.}\ }\textbf {\bibinfo {volume} {12}},\ \bibinfo {pages} {1012} (\bibinfo {year} {2020})}\BibitemShut {NoStop}%
\bibitem [{\citenamefont {Boaron}\ \emph {et~al.}(2018)\citenamefont {Boaron}, \citenamefont {Boso}, \citenamefont {Rusca}, \citenamefont {Vulliez}, \citenamefont {Autebert}, \citenamefont {Caloz}, \citenamefont {Perrenoud}, \citenamefont {Gras}, \citenamefont {Bussi\`eres}, \citenamefont {Li}, \citenamefont {Nolan}, \citenamefont {Martin},\ and\ \citenamefont {Zbinden}}]{PhysRevLett.121.190502}%
  \BibitemOpen
  \bibfield  {author} {\bibinfo {author} {\bibfnamefont {A.}~\bibnamefont {Boaron}}, \bibinfo {author} {\bibfnamefont {G.}~\bibnamefont {Boso}}, \bibinfo {author} {\bibfnamefont {D.}~\bibnamefont {Rusca}}, \bibinfo {author} {\bibfnamefont {C.}~\bibnamefont {Vulliez}}, \bibinfo {author} {\bibfnamefont {C.}~\bibnamefont {Autebert}}, \bibinfo {author} {\bibfnamefont {M.}~\bibnamefont {Caloz}}, \bibinfo {author} {\bibfnamefont {M.}~\bibnamefont {Perrenoud}}, \bibinfo {author} {\bibfnamefont {G.}~\bibnamefont {Gras}}, \bibinfo {author} {\bibfnamefont {F.}~\bibnamefont {Bussi\`eres}}, \bibinfo {author} {\bibfnamefont {M.-J.}\ \bibnamefont {Li}}, \bibinfo {author} {\bibfnamefont {D.}~\bibnamefont {Nolan}}, \bibinfo {author} {\bibfnamefont {A.}~\bibnamefont {Martin}},\ and\ \bibinfo {author} {\bibfnamefont {H.}~\bibnamefont {Zbinden}},\ }\bibfield  {title} {\bibinfo {title} {Secure quantum key distribution over 421 km of optical fiber},\ }\href {https://doi.org/10.1103/PhysRevLett.121.190502} {\bibfield  {journal}
  {\bibinfo  {journal} {Phys. Rev. Lett.}\ }\textbf {\bibinfo {volume} {121}},\ \bibinfo {pages} {190502} (\bibinfo {year} {2018})}\BibitemShut {NoStop}%
\bibitem [{\citenamefont {Dynes}\ \emph {et~al.}(2019)\citenamefont {Dynes}, \citenamefont {Wonfor}, \citenamefont {Tam}, \citenamefont {Sharpe}, \citenamefont {Takahashi}, \citenamefont {Lucamarini}, \citenamefont {Plews}, \citenamefont {Yuan}, \citenamefont {Dixon}, \citenamefont {Cho} \emph {et~al.}}]{dynes2019cambridge}%
  \BibitemOpen
  \bibfield  {author} {\bibinfo {author} {\bibfnamefont {J.}~\bibnamefont {Dynes}}, \bibinfo {author} {\bibfnamefont {A.}~\bibnamefont {Wonfor}}, \bibinfo {author} {\bibfnamefont {W.-S.}\ \bibnamefont {Tam}}, \bibinfo {author} {\bibfnamefont {A.}~\bibnamefont {Sharpe}}, \bibinfo {author} {\bibfnamefont {R.}~\bibnamefont {Takahashi}}, \bibinfo {author} {\bibfnamefont {M.}~\bibnamefont {Lucamarini}}, \bibinfo {author} {\bibfnamefont {A.}~\bibnamefont {Plews}}, \bibinfo {author} {\bibfnamefont {Z.}~\bibnamefont {Yuan}}, \bibinfo {author} {\bibfnamefont {A.}~\bibnamefont {Dixon}}, \bibinfo {author} {\bibfnamefont {J.}~\bibnamefont {Cho}}, \emph {et~al.},\ }\bibfield  {title} {\bibinfo {title} {Cambridge quantum network},\ }\href {https://doi.org/10.1038/s41534-019-0221-4} {\bibfield  {journal} {\bibinfo  {journal} {npj Quantum Inf.}\ }\textbf {\bibinfo {volume} {5}},\ \bibinfo {pages} {101} (\bibinfo {year} {2019})}\BibitemShut {NoStop}%
\bibitem [{\citenamefont {Ribezzo}\ \emph {et~al.}(2022)\citenamefont {Ribezzo}, \citenamefont {Zahidy}, \citenamefont {Vagniluca}, \citenamefont {Biagi}, \citenamefont {Francesconi}, \citenamefont {Occhipinti}, \citenamefont {Oxenløwe} \emph {et~al.}}]{ribezzo2023deploying}%
  \BibitemOpen
  \bibfield  {author} {\bibinfo {author} {\bibfnamefont {D.}~\bibnamefont {Ribezzo}}, \bibinfo {author} {\bibfnamefont {M.}~\bibnamefont {Zahidy}}, \bibinfo {author} {\bibfnamefont {I.}~\bibnamefont {Vagniluca}}, \bibinfo {author} {\bibfnamefont {N.}~\bibnamefont {Biagi}}, \bibinfo {author} {\bibfnamefont {S.}~\bibnamefont {Francesconi}}, \bibinfo {author} {\bibfnamefont {T.}~\bibnamefont {Occhipinti}}, \bibinfo {author} {\bibfnamefont {L.~K.}\ \bibnamefont {Oxenløwe}}, \emph {et~al.},\ }\bibfield  {title} {\bibinfo {title} {Deploying an inter-european quantum network},\ }\href {https://doi.org/https://doi.org/10.1002/qute.202200061} {\bibfield  {journal} {\bibinfo  {journal} {Adv. Quantum Technol.}\ }\textbf {\bibinfo {volume} {6}},\ \bibinfo {pages} {2200061} (\bibinfo {year} {2022})}\BibitemShut {NoStop}%
\bibitem [{\citenamefont {Martin}\ \emph {et~al.}(2024)\citenamefont {Martin}, \citenamefont {Brito}, \citenamefont {Ort{\'\i}z}, \citenamefont {Mendez}, \citenamefont {Buruaga}, \citenamefont {Vicente}, \citenamefont {Sebastian-Lombrana}, \citenamefont {Rincon}, \citenamefont {Perez}, \citenamefont {Sanchez} \emph {et~al.}}]{martin2024madqci}%
  \BibitemOpen
  \bibfield  {author} {\bibinfo {author} {\bibfnamefont {V.}~\bibnamefont {Martin}}, \bibinfo {author} {\bibfnamefont {J.~P.}\ \bibnamefont {Brito}}, \bibinfo {author} {\bibfnamefont {L.}~\bibnamefont {Ort{\'\i}z}}, \bibinfo {author} {\bibfnamefont {R.}~\bibnamefont {Mendez}}, \bibinfo {author} {\bibfnamefont {J.}~\bibnamefont {Buruaga}}, \bibinfo {author} {\bibfnamefont {R.}~\bibnamefont {Vicente}}, \bibinfo {author} {\bibfnamefont {A.}~\bibnamefont {Sebastian-Lombrana}}, \bibinfo {author} {\bibfnamefont {D.}~\bibnamefont {Rincon}}, \bibinfo {author} {\bibfnamefont {F.}~\bibnamefont {Perez}}, \bibinfo {author} {\bibfnamefont {C.}~\bibnamefont {Sanchez}}, \emph {et~al.},\ }\bibfield  {title} {\bibinfo {title} {Madqci: a heterogeneous and scalable sdn-qkd network deployed in production facilities},\ }\href {https://doi.org/10.1038/s41534-024-00873-2} {\bibfield  {journal} {\bibinfo  {journal} {npj Quantum Inf.}\ }\textbf {\bibinfo {volume} {10}},\ \bibinfo {pages} {80} (\bibinfo {year} {2024})}\BibitemShut
  {NoStop}%
\bibitem [{\citenamefont {Avesani}\ \emph {et~al.}(2021)\citenamefont {Avesani}, \citenamefont {Calderaro}, \citenamefont {Foletto}, \citenamefont {Agnesi}, \citenamefont {Picciariello}, \citenamefont {Santagiustina}, \citenamefont {Scriminich}, \citenamefont {Stanco}, \citenamefont {Vedovato}, \citenamefont {Zahidy}, \citenamefont {Vallone},\ and\ \citenamefont {Villoresi}}]{Avesani:21}%
  \BibitemOpen
  \bibfield  {author} {\bibinfo {author} {\bibfnamefont {M.}~\bibnamefont {Avesani}}, \bibinfo {author} {\bibfnamefont {L.}~\bibnamefont {Calderaro}}, \bibinfo {author} {\bibfnamefont {G.}~\bibnamefont {Foletto}}, \bibinfo {author} {\bibfnamefont {C.}~\bibnamefont {Agnesi}}, \bibinfo {author} {\bibfnamefont {F.}~\bibnamefont {Picciariello}}, \bibinfo {author} {\bibfnamefont {F.~B.~L.}\ \bibnamefont {Santagiustina}}, \bibinfo {author} {\bibfnamefont {A.}~\bibnamefont {Scriminich}}, \bibinfo {author} {\bibfnamefont {A.}~\bibnamefont {Stanco}}, \bibinfo {author} {\bibfnamefont {F.}~\bibnamefont {Vedovato}}, \bibinfo {author} {\bibfnamefont {M.}~\bibnamefont {Zahidy}}, \bibinfo {author} {\bibfnamefont {G.}~\bibnamefont {Vallone}},\ and\ \bibinfo {author} {\bibfnamefont {P.}~\bibnamefont {Villoresi}},\ }\bibfield  {title} {\bibinfo {title} {Resource-effective quantum key distribution: a field trial in {Padua} city center},\ }\href {https://doi.org/10.1364/OL.422890} {\bibfield  {journal} {\bibinfo  {journal} {Opt.
  Lett.}\ }\textbf {\bibinfo {volume} {46}},\ \bibinfo {pages} {2848} (\bibinfo {year} {2021})}\BibitemShut {NoStop}%
\bibitem [{\citenamefont {Agnesi}\ \emph {et~al.}(2024)\citenamefont {Agnesi}, \citenamefont {Giacomin}, \citenamefont {Sartorato}, \citenamefont {Artuso}, \citenamefont {Vallone},\ and\ \citenamefont {Villoresi}}]{Telebit}%
  \BibitemOpen
  \bibfield  {author} {\bibinfo {author} {\bibfnamefont {C.}~\bibnamefont {Agnesi}}, \bibinfo {author} {\bibfnamefont {M.}~\bibnamefont {Giacomin}}, \bibinfo {author} {\bibfnamefont {D.}~\bibnamefont {Sartorato}}, \bibinfo {author} {\bibfnamefont {S.}~\bibnamefont {Artuso}}, \bibinfo {author} {\bibfnamefont {G.}~\bibnamefont {Vallone}},\ and\ \bibinfo {author} {\bibfnamefont {P.}~\bibnamefont {Villoresi}},\ }\bibfield  {title} {\bibinfo {title} {In-field comparison between {G.652} and {G.655} optical fibres for polarisation-based quantum key distribution},\ }\href {https://doi.org/https://doi.org/10.1049/qtc2.12095} {\bibfield  {journal} {\bibinfo  {journal} {IET Quantum Comm.}\ }\textbf {\bibinfo {volume} {5}},\ \bibinfo {pages} {567} (\bibinfo {year} {2024})}\BibitemShut {NoStop}%
\bibitem [{\citenamefont {Li}\ \emph {et~al.}(2018)\citenamefont {Li}, \citenamefont {Gao}, \citenamefont {Li}, \citenamefont {Xue}, \citenamefont {Wang}, \citenamefont {Lu}, \citenamefont {Xiang}, \citenamefont {Zhao}, \citenamefont {Yan}, \citenamefont {Chen}, \citenamefont {Yu},\ and\ \citenamefont {Liu}}]{Li:18}%
  \BibitemOpen
  \bibfield  {author} {\bibinfo {author} {\bibfnamefont {D.-D.}\ \bibnamefont {Li}}, \bibinfo {author} {\bibfnamefont {S.}~\bibnamefont {Gao}}, \bibinfo {author} {\bibfnamefont {G.-C.}\ \bibnamefont {Li}}, \bibinfo {author} {\bibfnamefont {L.}~\bibnamefont {Xue}}, \bibinfo {author} {\bibfnamefont {L.-W.}\ \bibnamefont {Wang}}, \bibinfo {author} {\bibfnamefont {C.-B.}\ \bibnamefont {Lu}}, \bibinfo {author} {\bibfnamefont {Y.}~\bibnamefont {Xiang}}, \bibinfo {author} {\bibfnamefont {Z.-Y.}\ \bibnamefont {Zhao}}, \bibinfo {author} {\bibfnamefont {L.-C.}\ \bibnamefont {Yan}}, \bibinfo {author} {\bibfnamefont {Z.-Y.}\ \bibnamefont {Chen}}, \bibinfo {author} {\bibfnamefont {G.}~\bibnamefont {Yu}},\ and\ \bibinfo {author} {\bibfnamefont {J.-H.}\ \bibnamefont {Liu}},\ }\bibfield  {title} {\bibinfo {title} {Field implementation of long-distance quantum key distribution over aerial fiber with fast polarization feedback},\ }\href {https://doi.org/10.1364/OE.26.022793} {\bibfield  {journal} {\bibinfo  {journal} {Opt.
  Express}\ }\textbf {\bibinfo {volume} {26}},\ \bibinfo {pages} {22793} (\bibinfo {year} {2018})}\BibitemShut {NoStop}%
\bibitem [{\citenamefont {Islam}\ \emph {et~al.}(2017)\citenamefont {Islam}, \citenamefont {Lim}, \citenamefont {Cahall}, \citenamefont {Kim},\ and\ \citenamefont {Gauthier}}]{doi:10.1126/sciadv.1701491}%
  \BibitemOpen
  \bibfield  {author} {\bibinfo {author} {\bibfnamefont {N.~T.}\ \bibnamefont {Islam}}, \bibinfo {author} {\bibfnamefont {C.~C.~W.}\ \bibnamefont {Lim}}, \bibinfo {author} {\bibfnamefont {C.}~\bibnamefont {Cahall}}, \bibinfo {author} {\bibfnamefont {J.}~\bibnamefont {Kim}},\ and\ \bibinfo {author} {\bibfnamefont {D.~J.}\ \bibnamefont {Gauthier}},\ }\bibfield  {title} {\bibinfo {title} {Provably secure and high-rate quantum key distribution with time-bin qudits},\ }\href {https://doi.org/10.1126/sciadv.1701491} {\bibfield  {journal} {\bibinfo  {journal} {Sci. Adv.}\ }\textbf {\bibinfo {volume} {3}},\ \bibinfo {pages} {e1701491} (\bibinfo {year} {2017})}\BibitemShut {NoStop}%
\bibitem [{\citenamefont {Ecker}\ \emph {et~al.}(2019)\citenamefont {Ecker}, \citenamefont {Bouchard}, \citenamefont {Bulla}, \citenamefont {Brandt}, \citenamefont {Kohout}, \citenamefont {Steinlechner}, \citenamefont {Fickler}, \citenamefont {Malik}, \citenamefont {Guryanova}, \citenamefont {Ursin},\ and\ \citenamefont {Huber}}]{PhysRevX.9.041042}%
  \BibitemOpen
  \bibfield  {author} {\bibinfo {author} {\bibfnamefont {S.}~\bibnamefont {Ecker}}, \bibinfo {author} {\bibfnamefont {F.}~\bibnamefont {Bouchard}}, \bibinfo {author} {\bibfnamefont {L.}~\bibnamefont {Bulla}}, \bibinfo {author} {\bibfnamefont {F.}~\bibnamefont {Brandt}}, \bibinfo {author} {\bibfnamefont {O.}~\bibnamefont {Kohout}}, \bibinfo {author} {\bibfnamefont {F.}~\bibnamefont {Steinlechner}}, \bibinfo {author} {\bibfnamefont {R.}~\bibnamefont {Fickler}}, \bibinfo {author} {\bibfnamefont {M.}~\bibnamefont {Malik}}, \bibinfo {author} {\bibfnamefont {Y.}~\bibnamefont {Guryanova}}, \bibinfo {author} {\bibfnamefont {R.}~\bibnamefont {Ursin}},\ and\ \bibinfo {author} {\bibfnamefont {M.}~\bibnamefont {Huber}},\ }\bibfield  {title} {\bibinfo {title} {Overcoming noise in entanglement distribution},\ }\href {https://doi.org/10.1103/PhysRevX.9.041042} {\bibfield  {journal} {\bibinfo  {journal} {Phys. Rev. X}\ }\textbf {\bibinfo {volume} {9}},\ \bibinfo {pages} {041042} (\bibinfo {year} {2019})}\BibitemShut
  {NoStop}%
\bibitem [{\citenamefont {Sheridan}\ and\ \citenamefont {Scarani}(2010)}]{PhysRevA.82.030301}%
  \BibitemOpen
  \bibfield  {author} {\bibinfo {author} {\bibfnamefont {L.}~\bibnamefont {Sheridan}}\ and\ \bibinfo {author} {\bibfnamefont {V.}~\bibnamefont {Scarani}},\ }\bibfield  {title} {\bibinfo {title} {Security proof for quantum key distribution using qudit systems},\ }\href {https://doi.org/10.1103/PhysRevA.82.030301} {\bibfield  {journal} {\bibinfo  {journal} {Phys. Rev. A}\ }\textbf {\bibinfo {volume} {82}},\ \bibinfo {pages} {030301} (\bibinfo {year} {2010})}\BibitemShut {NoStop}%
\bibitem [{\citenamefont {Kaszlikowski}\ \emph {et~al.}(2000)\citenamefont {Kaszlikowski}, \citenamefont {Gnaci\ifmmode~\acute{n}\else \'{n}\fi{}ski}, \citenamefont {\ifmmode~\dot{Z}\else \.{Z}\fi{}ukowski}, \citenamefont {Miklaszewski},\ and\ \citenamefont {Zeilinger}}]{PhysRevLett.85.4418}%
  \BibitemOpen
  \bibfield  {author} {\bibinfo {author} {\bibfnamefont {D.}~\bibnamefont {Kaszlikowski}}, \bibinfo {author} {\bibfnamefont {P.}~\bibnamefont {Gnaci\ifmmode~\acute{n}\else \'{n}\fi{}ski}}, \bibinfo {author} {\bibfnamefont {M.}~\bibnamefont {\ifmmode~\dot{Z}\else \.{Z}\fi{}ukowski}}, \bibinfo {author} {\bibfnamefont {W.}~\bibnamefont {Miklaszewski}},\ and\ \bibinfo {author} {\bibfnamefont {A.}~\bibnamefont {Zeilinger}},\ }\bibfield  {title} {\bibinfo {title} {Violations of local realism by two entangled $\mathit{N}$-dimensional systems are stronger than for two qubits},\ }\href {https://doi.org/10.1103/PhysRevLett.85.4418} {\bibfield  {journal} {\bibinfo  {journal} {Phys. Rev. Lett.}\ }\textbf {\bibinfo {volume} {85}},\ \bibinfo {pages} {4418} (\bibinfo {year} {2000})}\BibitemShut {NoStop}%
\bibitem [{\citenamefont {Marcikic}\ \emph {et~al.}(2002)\citenamefont {Marcikic}, \citenamefont {de~Riedmatten}, \citenamefont {Tittel}, \citenamefont {Scarani}, \citenamefont {Zbinden},\ and\ \citenamefont {Gisin}}]{PhysRevA.66.062308}%
  \BibitemOpen
  \bibfield  {author} {\bibinfo {author} {\bibfnamefont {I.}~\bibnamefont {Marcikic}}, \bibinfo {author} {\bibfnamefont {H.}~\bibnamefont {de~Riedmatten}}, \bibinfo {author} {\bibfnamefont {W.}~\bibnamefont {Tittel}}, \bibinfo {author} {\bibfnamefont {V.}~\bibnamefont {Scarani}}, \bibinfo {author} {\bibfnamefont {H.}~\bibnamefont {Zbinden}},\ and\ \bibinfo {author} {\bibfnamefont {N.}~\bibnamefont {Gisin}},\ }\bibfield  {title} {\bibinfo {title} {Time-bin entangled qubits for quantum communication created by femtosecond pulses},\ }\href {https://doi.org/10.1103/PhysRevA.66.062308} {\bibfield  {journal} {\bibinfo  {journal} {Phys. Rev. A}\ }\textbf {\bibinfo {volume} {66}},\ \bibinfo {pages} {062308} (\bibinfo {year} {2002})}\BibitemShut {NoStop}%
\bibitem [{\citenamefont {Takesue}\ and\ \citenamefont {Noguchi}(2009)}]{Takesue:09}%
  \BibitemOpen
  \bibfield  {author} {\bibinfo {author} {\bibfnamefont {H.}~\bibnamefont {Takesue}}\ and\ \bibinfo {author} {\bibfnamefont {Y.}~\bibnamefont {Noguchi}},\ }\bibfield  {title} {\bibinfo {title} {Implementation of quantum state tomography for time-bin entangled photon pairs},\ }\href {https://doi.org/10.1364/OE.17.010976} {\bibfield  {journal} {\bibinfo  {journal} {Opt. Express}\ }\textbf {\bibinfo {volume} {17}},\ \bibinfo {pages} {10976} (\bibinfo {year} {2009})}\BibitemShut {NoStop}%
\bibitem [{\citenamefont {Vagniluca}\ \emph {et~al.}(2020)\citenamefont {Vagniluca}, \citenamefont {Da~Lio}, \citenamefont {Rusca}, \citenamefont {Cozzolino}, \citenamefont {Ding}, \citenamefont {Zbinden}, \citenamefont {Zavatta}, \citenamefont {Oxenl\o{}we},\ and\ \citenamefont {Bacco}}]{vagniluca2020efficient}%
  \BibitemOpen
  \bibfield  {author} {\bibinfo {author} {\bibfnamefont {I.}~\bibnamefont {Vagniluca}}, \bibinfo {author} {\bibfnamefont {B.}~\bibnamefont {Da~Lio}}, \bibinfo {author} {\bibfnamefont {D.}~\bibnamefont {Rusca}}, \bibinfo {author} {\bibfnamefont {D.}~\bibnamefont {Cozzolino}}, \bibinfo {author} {\bibfnamefont {Y.}~\bibnamefont {Ding}}, \bibinfo {author} {\bibfnamefont {H.}~\bibnamefont {Zbinden}}, \bibinfo {author} {\bibfnamefont {A.}~\bibnamefont {Zavatta}}, \bibinfo {author} {\bibfnamefont {L.~K.}\ \bibnamefont {Oxenl\o{}we}},\ and\ \bibinfo {author} {\bibfnamefont {D.}~\bibnamefont {Bacco}},\ }\bibfield  {title} {\bibinfo {title} {Efficient time-bin encoding for practical high-dimensional quantum key distribution},\ }\href {https://doi.org/10.1103/PhysRevApplied.14.014051} {\bibfield  {journal} {\bibinfo  {journal} {Phys. Rev. Appl.}\ }\textbf {\bibinfo {volume} {14}},\ \bibinfo {pages} {014051} (\bibinfo {year} {2020})}\BibitemShut {NoStop}%
\bibitem [{\citenamefont {Iskander}\ \emph {et~al.}(2019)\citenamefont {Iskander}, \citenamefont {Sinclair}, \citenamefont {Peña}, \citenamefont {Xie},\ and\ \citenamefont {Spiropulu}}]{Iskander2019}%
  \BibitemOpen
  \bibfield  {author} {\bibinfo {author} {\bibfnamefont {G.}~\bibnamefont {Iskander}}, \bibinfo {author} {\bibfnamefont {N.}~\bibnamefont {Sinclair}}, \bibinfo {author} {\bibfnamefont {C.}~\bibnamefont {Peña}}, \bibinfo {author} {\bibfnamefont {S.}~\bibnamefont {Xie}},\ and\ \bibinfo {author} {\bibfnamefont {M.}~\bibnamefont {Spiropulu}},\ }\bibfield  {title} {\bibinfo {title} {Stabilization of an electro-optic modulator for quantum communication using a low-cost microcontroller},\ }\href {https://curj.caltech.edu/2019/09/03/stabilization-of-an-electro-optic-modulator-for-quantum-communication-using-a-low-cost-microcontroller/} {\bibfield  {journal} {\bibinfo  {journal} {Caltech Undergraduate Res. J.}\ } (\bibinfo {year} {2019})}\BibitemShut {NoStop}%
\bibitem [{\citenamefont {Kim}\ \emph {et~al.}(2024)\citenamefont {Kim}, \citenamefont {Park}, \citenamefont {Kim}, \citenamefont {Kim}, \citenamefont {Kim}, \citenamefont {Park}, \citenamefont {Moon}, \citenamefont {Kwak}, \citenamefont {Kim},\ and\ \citenamefont {Ju}}]{Kim2024}%
  \BibitemOpen
  \bibfield  {author} {\bibinfo {author} {\bibfnamefont {J.}~\bibnamefont {Kim}}, \bibinfo {author} {\bibfnamefont {J.}~\bibnamefont {Park}}, \bibinfo {author} {\bibfnamefont {H.-S.}\ \bibnamefont {Kim}}, \bibinfo {author} {\bibfnamefont {G.}~\bibnamefont {Kim}}, \bibinfo {author} {\bibfnamefont {J.~T.}\ \bibnamefont {Kim}}, \bibinfo {author} {\bibfnamefont {J.}~\bibnamefont {Park}}, \bibinfo {author} {\bibfnamefont {K.}~\bibnamefont {Moon}}, \bibinfo {author} {\bibfnamefont {S.-C.}\ \bibnamefont {Kwak}}, \bibinfo {author} {\bibfnamefont {M.-s.}\ \bibnamefont {Kim}},\ and\ \bibinfo {author} {\bibfnamefont {J.~J.}\ \bibnamefont {Ju}},\ }\bibfield  {title} {\bibinfo {title} {Fully controllable time-bin entangled states distributed over 100-km single-mode fibers},\ }\href {http://dx.doi.org/10.1140/epjqt/s40507-024-00267-5} {\bibfield  {journal} {\bibinfo  {journal} {EPJ Quantum Technol.}\ }\textbf {\bibinfo {volume} {11}},\ \bibinfo {pages} {53} (\bibinfo {year} {2024})}\BibitemShut {NoStop}%
\bibitem [{\citenamefont {Tang}\ \emph {et~al.}(2013)\citenamefont {Tang}, \citenamefont {Yin}, \citenamefont {Ma}, \citenamefont {Fung}, \citenamefont {Liu}, \citenamefont {Yong}, \citenamefont {Chen}, \citenamefont {Peng}, \citenamefont {Chen},\ and\ \citenamefont {Pan}}]{PhysRevA.88.022308}%
  \BibitemOpen
  \bibfield  {author} {\bibinfo {author} {\bibfnamefont {Y.-L.}\ \bibnamefont {Tang}}, \bibinfo {author} {\bibfnamefont {H.-L.}\ \bibnamefont {Yin}}, \bibinfo {author} {\bibfnamefont {X.}~\bibnamefont {Ma}}, \bibinfo {author} {\bibfnamefont {C.-H.~F.}\ \bibnamefont {Fung}}, \bibinfo {author} {\bibfnamefont {Y.}~\bibnamefont {Liu}}, \bibinfo {author} {\bibfnamefont {H.-L.}\ \bibnamefont {Yong}}, \bibinfo {author} {\bibfnamefont {T.-Y.}\ \bibnamefont {Chen}}, \bibinfo {author} {\bibfnamefont {C.-Z.}\ \bibnamefont {Peng}}, \bibinfo {author} {\bibfnamefont {Z.-B.}\ \bibnamefont {Chen}},\ and\ \bibinfo {author} {\bibfnamefont {J.-W.}\ \bibnamefont {Pan}},\ }\bibfield  {title} {\bibinfo {title} {Source attack of decoy-state quantum key distribution using phase information},\ }\href {https://doi.org/10.1103/PhysRevA.88.022308} {\bibfield  {journal} {\bibinfo  {journal} {Phys. Rev. A}\ }\textbf {\bibinfo {volume} {88}},\ \bibinfo {pages} {022308} (\bibinfo {year} {2013})}\BibitemShut {NoStop}%
\bibitem [{\citenamefont {Ac\'{\i}n}\ \emph {et~al.}(2012)\citenamefont {Ac\'{\i}n}, \citenamefont {Massar},\ and\ \citenamefont {Pironio}}]{PhysRevLett.108.100402}%
  \BibitemOpen
  \bibfield  {author} {\bibinfo {author} {\bibfnamefont {A.}~\bibnamefont {Ac\'{\i}n}}, \bibinfo {author} {\bibfnamefont {S.}~\bibnamefont {Massar}},\ and\ \bibinfo {author} {\bibfnamefont {S.}~\bibnamefont {Pironio}},\ }\bibfield  {title} {\bibinfo {title} {Randomness versus nonlocality and entanglement},\ }\href {https://doi.org/10.1103/PhysRevLett.108.100402} {\bibfield  {journal} {\bibinfo  {journal} {Phys. Rev. Lett.}\ }\textbf {\bibinfo {volume} {108}},\ \bibinfo {pages} {100402} (\bibinfo {year} {2012})}\BibitemShut {NoStop}%
\bibitem [{\citenamefont {Roberts}\ \emph {et~al.}(2018)\citenamefont {Roberts}, \citenamefont {Pittaluga}, \citenamefont {Minder}, \citenamefont {Lucamarini}, \citenamefont {Dynes}, \citenamefont {Yuan},\ and\ \citenamefont {Shields}}]{Roberts:18}%
  \BibitemOpen
  \bibfield  {author} {\bibinfo {author} {\bibfnamefont {G.~L.}\ \bibnamefont {Roberts}}, \bibinfo {author} {\bibfnamefont {M.}~\bibnamefont {Pittaluga}}, \bibinfo {author} {\bibfnamefont {M.}~\bibnamefont {Minder}}, \bibinfo {author} {\bibfnamefont {M.}~\bibnamefont {Lucamarini}}, \bibinfo {author} {\bibfnamefont {J.~F.}\ \bibnamefont {Dynes}}, \bibinfo {author} {\bibfnamefont {Z.~L.}\ \bibnamefont {Yuan}},\ and\ \bibinfo {author} {\bibfnamefont {A.~J.}\ \bibnamefont {Shields}},\ }\bibfield  {title} {\bibinfo {title} {Patterning-effect mitigating intensity modulator for secure decoy-state quantum key distribution},\ }\href {https://doi.org/10.1364/OL.43.005110} {\bibfield  {journal} {\bibinfo  {journal} {Opt. Lett.}\ }\textbf {\bibinfo {volume} {43}},\ \bibinfo {pages} {5110} (\bibinfo {year} {2018})}\BibitemShut {NoStop}%
\bibitem [{\citenamefont {Avesani}\ \emph {et~al.}(2020)\citenamefont {Avesani}, \citenamefont {Agnesi}, \citenamefont {Stanco}, \citenamefont {Vallone},\ and\ \citenamefont {Villoresi}}]{iPognac}%
  \BibitemOpen
  \bibfield  {author} {\bibinfo {author} {\bibfnamefont {M.}~\bibnamefont {Avesani}}, \bibinfo {author} {\bibfnamefont {C.}~\bibnamefont {Agnesi}}, \bibinfo {author} {\bibfnamefont {A.}~\bibnamefont {Stanco}}, \bibinfo {author} {\bibfnamefont {G.}~\bibnamefont {Vallone}},\ and\ \bibinfo {author} {\bibfnamefont {P.}~\bibnamefont {Villoresi}},\ }\bibfield  {title} {\bibinfo {title} {Stable, low-error, and calibration-free polarization encoder for free-space quantum communication},\ }\href {https://doi.org/10.1364/OL.396412} {\bibfield  {journal} {\bibinfo  {journal} {Opt. Lett.}\ }\textbf {\bibinfo {volume} {45}},\ \bibinfo {pages} {4706} (\bibinfo {year} {2020})}\BibitemShut {NoStop}%
\bibitem [{\citenamefont {Paraïso}\ \emph {et~al.}(2021)\citenamefont {Paraïso}, \citenamefont {Woodward}, \citenamefont {Marangon}, \citenamefont {Lovic}, \citenamefont {Yuan},\ and\ \citenamefont {Shields}}]{Paraso2021}%
  \BibitemOpen
  \bibfield  {author} {\bibinfo {author} {\bibfnamefont {T.~K.}\ \bibnamefont {Paraïso}}, \bibinfo {author} {\bibfnamefont {R.~I.}\ \bibnamefont {Woodward}}, \bibinfo {author} {\bibfnamefont {D.~G.}\ \bibnamefont {Marangon}}, \bibinfo {author} {\bibfnamefont {V.}~\bibnamefont {Lovic}}, \bibinfo {author} {\bibfnamefont {Z.}~\bibnamefont {Yuan}},\ and\ \bibinfo {author} {\bibfnamefont {A.~J.}\ \bibnamefont {Shields}},\ }\bibfield  {title} {\bibinfo {title} {Advanced laser technology for quantum communications (tutorial review)},\ }\href {https://doi.org/https://doi.org/10.1002/qute.202100062} {\bibfield  {journal} {\bibinfo  {journal} {Adv. Quantum Technol.}\ }\textbf {\bibinfo {volume} {4}},\ \bibinfo {pages} {2100062} (\bibinfo {year} {2021})}\BibitemShut {NoStop}%
\bibitem [{\citenamefont {Rusca}\ \emph {et~al.}(2018{\natexlab{a}})\citenamefont {Rusca}, \citenamefont {Boaron}, \citenamefont {Curty}, \citenamefont {Martin},\ and\ \citenamefont {Zbinden}}]{rusca2018security}%
  \BibitemOpen
  \bibfield  {author} {\bibinfo {author} {\bibfnamefont {D.}~\bibnamefont {Rusca}}, \bibinfo {author} {\bibfnamefont {A.}~\bibnamefont {Boaron}}, \bibinfo {author} {\bibfnamefont {M.}~\bibnamefont {Curty}}, \bibinfo {author} {\bibfnamefont {A.}~\bibnamefont {Martin}},\ and\ \bibinfo {author} {\bibfnamefont {H.}~\bibnamefont {Zbinden}},\ }\bibfield  {title} {\bibinfo {title} {Security proof for a simplified bennett-brassard 1984 quantum-key-distribution protocol},\ }\href {https://doi.org/10.1103/PhysRevA.98.052336} {\bibfield  {journal} {\bibinfo  {journal} {Phys. Rev. A}\ }\textbf {\bibinfo {volume} {98}},\ \bibinfo {pages} {052336} (\bibinfo {year} {2018}{\natexlab{a}})}\BibitemShut {NoStop}%
\bibitem [{\citenamefont {Lu}\ \emph {et~al.}(2022)\citenamefont {Lu}, \citenamefont {Cao}, \citenamefont {Peng},\ and\ \citenamefont {Pan}}]{lu2022micius}%
  \BibitemOpen
  \bibfield  {author} {\bibinfo {author} {\bibfnamefont {C.-Y.}\ \bibnamefont {Lu}}, \bibinfo {author} {\bibfnamefont {Y.}~\bibnamefont {Cao}}, \bibinfo {author} {\bibfnamefont {C.-Z.}\ \bibnamefont {Peng}},\ and\ \bibinfo {author} {\bibfnamefont {J.-W.}\ \bibnamefont {Pan}},\ }\bibfield  {title} {\bibinfo {title} {Micius quantum experiments in space},\ }\href {https://doi.org/10.1103/RevModPhys.94.035001} {\bibfield  {journal} {\bibinfo  {journal} {Rev. Mod. Phys.}\ }\textbf {\bibinfo {volume} {94}},\ \bibinfo {pages} {035001} (\bibinfo {year} {2022})}\BibitemShut {NoStop}%
\bibitem [{\citenamefont {Li}\ \emph {et~al.}(2025)\citenamefont {Li}, \citenamefont {Cai}, \citenamefont {Ren}, \citenamefont {Wang}, \citenamefont {Yang}, \citenamefont {Zhang}, \citenamefont {Wu}, \citenamefont {Chang}, \citenamefont {Wu}, \citenamefont {Jin} \emph {et~al.}}]{li2025microsatellite}%
  \BibitemOpen
  \bibfield  {author} {\bibinfo {author} {\bibfnamefont {Y.}~\bibnamefont {Li}}, \bibinfo {author} {\bibfnamefont {W.-Q.}\ \bibnamefont {Cai}}, \bibinfo {author} {\bibfnamefont {J.-G.}\ \bibnamefont {Ren}}, \bibinfo {author} {\bibfnamefont {C.-Z.}\ \bibnamefont {Wang}}, \bibinfo {author} {\bibfnamefont {M.}~\bibnamefont {Yang}}, \bibinfo {author} {\bibfnamefont {L.}~\bibnamefont {Zhang}}, \bibinfo {author} {\bibfnamefont {H.-Y.}\ \bibnamefont {Wu}}, \bibinfo {author} {\bibfnamefont {L.}~\bibnamefont {Chang}}, \bibinfo {author} {\bibfnamefont {J.-C.}\ \bibnamefont {Wu}}, \bibinfo {author} {\bibfnamefont {B.}~\bibnamefont {Jin}}, \emph {et~al.},\ }\bibfield  {title} {\bibinfo {title} {Microsatellite-based real-time quantum key distribution},\ }\href {https://doi.org/10.1038/s41586-025-08739-z} {\bibfield  {journal} {\bibinfo  {journal} {Nature}\ }\textbf {\bibinfo {volume} {640}},\ \bibinfo {pages} {47} (\bibinfo {year} {2025})}\BibitemShut {NoStop}%
\bibitem [{\citenamefont {\v{S}varc}\ \emph {et~al.}(2023)\citenamefont {\v{S}varc}, \citenamefont {Nov\'{a}kov\'{a}}, \citenamefont {Dudka},\ and\ \citenamefont {Je\v{z}ek}}]{vsvarc2023sub}%
  \BibitemOpen
  \bibfield  {author} {\bibinfo {author} {\bibfnamefont {V.}~\bibnamefont {\v{S}varc}}, \bibinfo {author} {\bibfnamefont {M.}~\bibnamefont {Nov\'{a}kov\'{a}}}, \bibinfo {author} {\bibfnamefont {M.}~\bibnamefont {Dudka}},\ and\ \bibinfo {author} {\bibfnamefont {M.}~\bibnamefont {Je\v{z}ek}},\ }\bibfield  {title} {\bibinfo {title} {Sub-0.1 degree phase locking of a single-photon interferometer},\ }\href {https://doi.org/10.1364/OE.480569} {\bibfield  {journal} {\bibinfo  {journal} {Opt. Express}\ }\textbf {\bibinfo {volume} {31}},\ \bibinfo {pages} {12562} (\bibinfo {year} {2023})}\BibitemShut {NoStop}%
\bibitem [{\citenamefont {Rusca}\ \emph {et~al.}(2018{\natexlab{b}})\citenamefont {Rusca}, \citenamefont {Boaron}, \citenamefont {Grünenfelder}, \citenamefont {Martin},\ and\ \citenamefont {Zbinden}}]{rusca2018finite}%
  \BibitemOpen
  \bibfield  {author} {\bibinfo {author} {\bibfnamefont {D.}~\bibnamefont {Rusca}}, \bibinfo {author} {\bibfnamefont {A.}~\bibnamefont {Boaron}}, \bibinfo {author} {\bibfnamefont {F.}~\bibnamefont {Grünenfelder}}, \bibinfo {author} {\bibfnamefont {A.}~\bibnamefont {Martin}},\ and\ \bibinfo {author} {\bibfnamefont {H.}~\bibnamefont {Zbinden}},\ }\bibfield  {title} {\bibinfo {title} {Finite-key analysis for the 1-decoy state qkd protocol},\ }\href {https://doi.org/10.1063/1.5023340} {\bibfield  {journal} {\bibinfo  {journal} {Appl. Phys. Lett.}\ }\textbf {\bibinfo {volume} {112}},\ \bibinfo {pages} {171104} (\bibinfo {year} {2018}{\natexlab{b}})}\BibitemShut {NoStop}%
\bibitem [{\citenamefont {Curr{\'a}s-Lorenzo}\ \emph {et~al.}(2023)\citenamefont {Curr{\'a}s-Lorenzo}, \citenamefont {Nahar}, \citenamefont {L{\"u}tkenhaus}, \citenamefont {Tamaki},\ and\ \citenamefont {Curty}}]{curras2023security}%
  \BibitemOpen
  \bibfield  {author} {\bibinfo {author} {\bibfnamefont {G.}~\bibnamefont {Curr{\'a}s-Lorenzo}}, \bibinfo {author} {\bibfnamefont {S.}~\bibnamefont {Nahar}}, \bibinfo {author} {\bibfnamefont {N.}~\bibnamefont {L{\"u}tkenhaus}}, \bibinfo {author} {\bibfnamefont {K.}~\bibnamefont {Tamaki}},\ and\ \bibinfo {author} {\bibfnamefont {M.}~\bibnamefont {Curty}},\ }\bibfield  {title} {\bibinfo {title} {Security of quantum key distribution with imperfect phase randomisation},\ }\href {https://doi.org/10.1088/2058-9565/ad141c} {\bibfield  {journal} {\bibinfo  {journal} {Quantum Sci. Technol.}\ }\textbf {\bibinfo {volume} {9}},\ \bibinfo {pages} {015025} (\bibinfo {year} {2023})}\BibitemShut {NoStop}%
\bibitem [{\citenamefont {Stanco}\ \emph {et~al.}(2022)\citenamefont {Stanco}, \citenamefont {Santagiustina}, \citenamefont {Calderaro}, \citenamefont {Avesani}, \citenamefont {Bertapelle}, \citenamefont {Dequal}, \citenamefont {Vallone},\ and\ \citenamefont {Villoresi}}]{9695406}%
  \BibitemOpen
  \bibfield  {author} {\bibinfo {author} {\bibfnamefont {A.}~\bibnamefont {Stanco}}, \bibinfo {author} {\bibfnamefont {F.~B.~L.}\ \bibnamefont {Santagiustina}}, \bibinfo {author} {\bibfnamefont {L.}~\bibnamefont {Calderaro}}, \bibinfo {author} {\bibfnamefont {M.}~\bibnamefont {Avesani}}, \bibinfo {author} {\bibfnamefont {T.}~\bibnamefont {Bertapelle}}, \bibinfo {author} {\bibfnamefont {D.}~\bibnamefont {Dequal}}, \bibinfo {author} {\bibfnamefont {G.}~\bibnamefont {Vallone}},\ and\ \bibinfo {author} {\bibfnamefont {P.}~\bibnamefont {Villoresi}},\ }\bibfield  {title} {\bibinfo {title} {Versatile and concurrent {FPGA}-based architecture for practical quantum communication systems},\ }\href {https://doi.org/10.1109/TQE.2022.3143997} {\bibfield  {journal} {\bibinfo  {journal} {IEEE Trans. Quantum Eng.}\ }\textbf {\bibinfo {volume} {3}},\ \bibinfo {pages} {1} (\bibinfo {year} {2022})}\BibitemShut {NoStop}%
\bibitem [{\citenamefont {Francesconi}\ \emph {et~al.}(2024)\citenamefont {Francesconi}, \citenamefont {De~Lazzari}, \citenamefont {Ribezzo}, \citenamefont {Vagniluca}, \citenamefont {Biagi}, \citenamefont {Occhipinti}, \citenamefont {Zavatta},\ and\ \citenamefont {Bacco}}]{francesconi2024scalable}%
  \BibitemOpen
  \bibfield  {author} {\bibinfo {author} {\bibfnamefont {S.}~\bibnamefont {Francesconi}}, \bibinfo {author} {\bibfnamefont {C.}~\bibnamefont {De~Lazzari}}, \bibinfo {author} {\bibfnamefont {D.}~\bibnamefont {Ribezzo}}, \bibinfo {author} {\bibfnamefont {I.}~\bibnamefont {Vagniluca}}, \bibinfo {author} {\bibfnamefont {N.}~\bibnamefont {Biagi}}, \bibinfo {author} {\bibfnamefont {T.}~\bibnamefont {Occhipinti}}, \bibinfo {author} {\bibfnamefont {A.}~\bibnamefont {Zavatta}},\ and\ \bibinfo {author} {\bibfnamefont {D.}~\bibnamefont {Bacco}},\ }\bibfield  {title} {\bibinfo {title} {Scalable implementation of temporal and phase encoding qkd with phase-randomized states},\ }\href {https://doi.org/https://doi.org/10.1002/qute.202300224} {\bibfield  {journal} {\bibinfo  {journal} {Adv Quantum Technol.}\ }\textbf {\bibinfo {volume} {7}},\ \bibinfo {pages} {2300224} (\bibinfo {year} {2024})}\BibitemShut {NoStop}%
\bibitem [{\citenamefont {Grünenfelder}\ \emph {et~al.}(2020)\citenamefont {Grünenfelder}, \citenamefont {Boaron}, \citenamefont {Rusca}, \citenamefont {Martin},\ and\ \citenamefont {Zbinden}}]{grunenfelder2020performance}%
  \BibitemOpen
  \bibfield  {author} {\bibinfo {author} {\bibfnamefont {F.}~\bibnamefont {Grünenfelder}}, \bibinfo {author} {\bibfnamefont {A.}~\bibnamefont {Boaron}}, \bibinfo {author} {\bibfnamefont {D.}~\bibnamefont {Rusca}}, \bibinfo {author} {\bibfnamefont {A.}~\bibnamefont {Martin}},\ and\ \bibinfo {author} {\bibfnamefont {H.}~\bibnamefont {Zbinden}},\ }\bibfield  {title} {\bibinfo {title} {Performance and security of 5 {GHz} repetition rate polarization-based quantum key distribution},\ }\href {https://doi.org/10.1063/5.0021468} {\bibfield  {journal} {\bibinfo  {journal} {Appl. Phys. Lett.}\ }\textbf {\bibinfo {volume} {117}},\ \bibinfo {pages} {144003} (\bibinfo {year} {2020})}\BibitemShut {NoStop}%
\bibitem [{\citenamefont {Thew}\ \emph {et~al.}(2002)\citenamefont {Thew}, \citenamefont {Nemoto}, \citenamefont {White},\ and\ \citenamefont {Munro}}]{PhysRevA.66.012303}%
  \BibitemOpen
  \bibfield  {author} {\bibinfo {author} {\bibfnamefont {R.~T.}\ \bibnamefont {Thew}}, \bibinfo {author} {\bibfnamefont {K.}~\bibnamefont {Nemoto}}, \bibinfo {author} {\bibfnamefont {A.~G.}\ \bibnamefont {White}},\ and\ \bibinfo {author} {\bibfnamefont {W.~J.}\ \bibnamefont {Munro}},\ }\bibfield  {title} {\bibinfo {title} {Qudit quantum-state tomography},\ }\href {https://doi.org/10.1103/PhysRevA.66.012303} {\bibfield  {journal} {\bibinfo  {journal} {Phys. Rev. A}\ }\textbf {\bibinfo {volume} {66}},\ \bibinfo {pages} {012303} (\bibinfo {year} {2002})}\BibitemShut {NoStop}%
\bibitem [{\citenamefont {Jogenfors}\ and\ \citenamefont {Larsson}(2014)}]{Jogenfors_2014}%
  \BibitemOpen
  \bibfield  {author} {\bibinfo {author} {\bibfnamefont {J.}~\bibnamefont {Jogenfors}}\ and\ \bibinfo {author} {\bibfnamefont {J.-{\r A}.}\ \bibnamefont {Larsson}},\ }\bibfield  {title} {\bibinfo {title} {Energy-time entanglement, elements of reality, and local realism},\ }\href {https://doi.org/10.1088/1751-8113/47/42/424032} {\bibfield  {journal} {\bibinfo  {journal} {J. Phys. A: Math. Theor.}\ }\textbf {\bibinfo {volume} {47}},\ \bibinfo {pages} {424032} (\bibinfo {year} {2014})}\BibitemShut {NoStop}%
\bibitem [{\citenamefont {Vedovato}\ \emph {et~al.}(2018)\citenamefont {Vedovato}, \citenamefont {Agnesi}, \citenamefont {Tomasin}, \citenamefont {Avesani}, \citenamefont {Larsson}, \citenamefont {Vallone},\ and\ \citenamefont {Villoresi}}]{PhysRevLett.121.190401}%
  \BibitemOpen
  \bibfield  {author} {\bibinfo {author} {\bibfnamefont {F.}~\bibnamefont {Vedovato}}, \bibinfo {author} {\bibfnamefont {C.}~\bibnamefont {Agnesi}}, \bibinfo {author} {\bibfnamefont {M.}~\bibnamefont {Tomasin}}, \bibinfo {author} {\bibfnamefont {M.}~\bibnamefont {Avesani}}, \bibinfo {author} {\bibfnamefont {J.-{\r A}.}\ \bibnamefont {Larsson}}, \bibinfo {author} {\bibfnamefont {G.}~\bibnamefont {Vallone}},\ and\ \bibinfo {author} {\bibfnamefont {P.}~\bibnamefont {Villoresi}},\ }\bibfield  {title} {\bibinfo {title} {Postselection-loophole-free bell violation with genuine time-bin entanglement},\ }\href {https://doi.org/10.1103/PhysRevLett.121.190401} {\bibfield  {journal} {\bibinfo  {journal} {Phys. Rev. Lett.}\ }\textbf {\bibinfo {volume} {121}},\ \bibinfo {pages} {190401} (\bibinfo {year} {2018})}\BibitemShut {NoStop}%
\bibitem [{\citenamefont {Santagiustina}\ \emph {et~al.}(2024)\citenamefont {Santagiustina}, \citenamefont {Agnesi}, \citenamefont {Alarc\'{o}n}, \citenamefont {Cabello}, \citenamefont {Xavier}, \citenamefont {Villoresi},\ and\ \citenamefont {Vallone}}]{Santagiustina:24}%
  \BibitemOpen
  \bibfield  {author} {\bibinfo {author} {\bibfnamefont {F.~B.~L.}\ \bibnamefont {Santagiustina}}, \bibinfo {author} {\bibfnamefont {C.}~\bibnamefont {Agnesi}}, \bibinfo {author} {\bibfnamefont {A.}~\bibnamefont {Alarc\'{o}n}}, \bibinfo {author} {\bibfnamefont {A.}~\bibnamefont {Cabello}}, \bibinfo {author} {\bibfnamefont {G.~B.}\ \bibnamefont {Xavier}}, \bibinfo {author} {\bibfnamefont {P.}~\bibnamefont {Villoresi}},\ and\ \bibinfo {author} {\bibfnamefont {G.}~\bibnamefont {Vallone}},\ }\bibfield  {title} {\bibinfo {title} {Experimental post-selection loophole-free time-bin and energy-time nonlocality with integrated photonics},\ }\href {https://doi.org/10.1364/OPTICA.499247} {\bibfield  {journal} {\bibinfo  {journal} {Optica}\ }\textbf {\bibinfo {volume} {11}},\ \bibinfo {pages} {498} (\bibinfo {year} {2024})}\BibitemShut {NoStop}%
\end{thebibliography}%

\section*{Acknowledgment}
K.V. acknowledges support from the European Union’s Horizon Europe Framework Programme under the Marie Sklodowska Curie Grant No. 956071, Project AppQInfo, and Grant  No. 101070168, Project HyperSpace.

M.R.B. acknowledges support from the European Union’s Horizon Europe Framework Programme under the Marie Sklodowska Curie Grant No. 101072637, Project Quantum-Safe Internet (QSI).

The authors would like to thank Davide G. Marangon for fruitful discussions.

\end{document}